\documentclass[iop,twocolappendix,numberedappendix]{emulateapj}
\usepackage{graphicx}
\usepackage{subfigure}
\usepackage{multirow}
\usepackage{amssymb}
\usepackage{natbib}
\usepackage{rotating}
\usepackage[bookmarks=false, colorlinks=true, citecolor=blue, backref=true]{hyperref}
\usepackage{apjfonts}

\shorttitle{Evolution of Star-Formation Enhancement in Galaxy Pairs}
\shortauthors{C.K. Xu et al.}

\newcommand{\lsim}{\, \lower2truept\hbox{${< \atop\hbox{\raise4truept\hbox{$\sim$}}}$}\,}
\newcommand{\gsim}{\, \lower2truept\hbox{${> \atop\hbox{\raise4truept\hbox{$\sim$}}}$}\,}

\begin{document}

\slugcomment{{\bf Draft 7}; \today}

\title{Cosmic Evolution of Star-Formation Enhancement in Close Major-Merger Galaxy Pairs Since $z = 1$}
\thanks{\textit{Herschel} is an ESA space observatory with science instruments provided by European-led Principal Investigator consortia and with important participation from NASA.}

% hermes_authors script, version DR1 2012-01-26
\author{C.K.~Xu\altaffilmark{1,2},
D.L.~Shupe\altaffilmark{1,2},
M.~B{\'e}thermin\altaffilmark{3,4},
H.~Aussel\altaffilmark{3},
S.~Berta\altaffilmark{5},
J.~Bock\altaffilmark{1,6},
C.~Bridge\altaffilmark{1},
A.~Conley\altaffilmark{7},
A.~Cooray\altaffilmark{8,1},
D.~Elbaz\altaffilmark{3},
A.~Franceschini\altaffilmark{9},
E.~Le Floc'h\altaffilmark{3},
N.~Lu\altaffilmark{1,2},
D.~Lutz\altaffilmark{5},
B.~Magnelli\altaffilmark{5},
G.~Marsden\altaffilmark{10},
S.J.~Oliver\altaffilmark{11},
F.~Pozzi\altaffilmark{12},
L.~Riguccini\altaffilmark{3},
B.~Schulz\altaffilmark{1,2},
N.~Scoville\altaffilmark{1},
M.~Vaccari\altaffilmark{9,13},
J.D.~Vieira\altaffilmark{1},
L.~Wang\altaffilmark{11},
M.~Zemcov\altaffilmark{1,6}}
\altaffiltext{1}{California Institute of Technology, 1200 E. California Blvd., Pasadena, CA 91125}
\altaffiltext{2}{Infrared Processing and Analysis Center, MS 100-22, California Institute of Technology, JPL, Pasadena, CA 91125}
\altaffiltext{3}{Laboratoire AIM-Paris-Saclay, CEA/DSM/Irfu - CNRS - Universit\'e Paris Diderot, CE-Saclay, pt courrier 131, F-91191 Gif-sur-Yvette, France}
\altaffiltext{4}{Institut d'Astrophysique Spatiale (IAS), b\^atiment 121, Universit\'e Paris-Sud 11 and CNRS (UMR 8617), 91405 Orsay, France}
\altaffiltext{5}{Max-Planck-Institut f\"ur Extraterrestrische Physik (MPE), Postfach 1312, 85741, Garching, Germany}
\altaffiltext{6}{Jet Propulsion Laboratory, 4800 Oak Grove Drive, Pasadena, CA 91109}
\altaffiltext{7}{Center for Astrophysics and Space Astronomy 389-UCB, University of Colorado, Boulder, CO 80309}
\altaffiltext{8}{Dept. of Physics \& Astronomy, University of California, Irvine, CA 92697}
\altaffiltext{9}{Dipartimento di Astronomia, Universit\`{a} di Padova, vicolo Osservatorio, 3, 35122 Padova, Italy}
\altaffiltext{10}{Department of Physics \& Astronomy, University of British Columbia, 6224 Agricultural Road, Vancouver, BC V6T~1Z1, Canada}
\altaffiltext{11}{Astronomy Centre, Dept. of Physics \& Astronomy, University of Sussex, Brighton BN1 9QH, UK}
\altaffiltext{12}{INAF-Osservatorio Astronomico di Roma, via di Frascati 33, 00040 Monte Porzio Catone, Italy}
\altaffiltext{13}{Astrophysics Group, Physics Department, University of the Western Cape, Private Bag X17, 7535, Bellville, Cape Town, South Africa}

%\received{Sept.~7, 2011}
%\accepted{Nov.~22, 2011}

\begin{abstract}
The infrared (IR) emission of ``$\rm M_*$ galaxies'' ($\rm 10^{10.4}
\leq M_{star} \leq 10^{11.0}\; M_\sun$) in galaxy pairs, derived using
data obtained in \textit{Herschel} (PEP/HerMES) and \textit{Spitzer}
(S-COSMOS) surveys, is compared to that of single disk galaxies in
well matched control samples to study the cosmic evolution of the
star-formation enhancement induced by galaxy-galaxy interaction.  Both
the mean IR SED and mean IR luminosity of star-forming galaxies (SFGs)
in SFG+SFG (S+S) pairs in the redshift bin of $\rm 0.6 < z < 1$ are
consistent with no star-formation enhancement.  SFGs in S+S pairs in a
lower redshift bin of $\rm 0.2 < z < 0.6$ show marginal evidence for a
weak star-formation enhancement. Together with the significant and
strong sSFR enhancement shown by SFGs in a local sample of S+S pairs
(obtained using previously published Spitzer observations), our
results reveal a trend for the star-formation enhancement in S+S pairs
to decrease with increasing redshift.  Between $z=0$ and $z= 1$, this
decline of interaction-induced star-formation enhancement occurs in
parallel with the dramatic increase (by a factor of $\sim 10$) of the
sSFR of single SFGs, both can be explained by the higher gas fraction
in higher z disks.  SFGs in mixed pairs (S+E pairs) do not show any
significant star-formation enhancement at any redshift.  The
difference between SFGs in S+S pairs and in S+E pairs suggests a
modulation of the sSFR by the inter-galactic medium (IGM) in the dark
matter halos (DMH) hosting these pairs.
\end{abstract}


\section{Introduction}
In some cosmological simulations \citep{Guiderdoni1998,
  Somerville2001, Baugh2005}, it is assumed that merger-induced
star-formation is the major (or even the dominant) contribution
to the high star-formation rate (SFR) at $z \sim 1$ -- 2, the epoch
when the SFR density in the universe peaks.  Results of early HST surveys
\citep{Driver1995, Glazebrook1995, Abraham1996, Brinchmann1998,
  LeFevre2000, Conselice2003} were indeed consistent with this
assumption. However, more recent observations \citep{Bell2005, Noeske2007a,
  Elbaz2007, Daddi2007, Lotz2008, Pannella2009, Rodighiero2011,
  Peng2010, Elbaz2011, Wuyts2011} have favored a scenario in which the
so-called ``main sequence'' (MS) normal star-forming galaxies (SFG),
continuously fueled by smooth accretion of cold intergalactic gas
(``cold flows'', \citealt{Dekel2009, Keres2009}), dominate the cosmic
SFR ever since z$\sim 4$ while the contribution from merger-induced
starbursts (the outliers in the SFR-$\rm M_{star}$ plot) plays only a
minor role (\citealt{Rodighiero2011}, but see \citealt{Bridge2007,
  Shi2009}).
 
On the other hand, it is possible that much of the SFR in the MS
galaxies could be associated with galaxy mergers.
Many recent studies on merger rate
evolution found that the
fraction of galaxies in mergers increases significantly from z=0 to
z=1 (\citealt{Kartaltepe2007, Conselice2009, Bridge2010, Xu2012},
and $\sim 10\%$ galaxies with $z\sim 1$ are in close major-merger
pairs (see \citealt{Lin2008, Lotz2008} for different results).
In the local universe, the
average SFR of merging galaxies (as found in optical/NIR selected
pairs) is only a factor of $\sim 2$ -- 3 of that of single disk
galaxies \citep{Kennicutt1987, Ellison2008, Xu2010}. Therefore most of
these galaxies belong to the MS population, while extreme starbursts
such as those in local ultra-luminous galaxies (ULIRGs) are very rare
\citep{Sanders1996}.  If mergers of higher redshifts have similar or
higher level of SFR enhancement, then $\gsim 30\%$ of the SFR
associated with MS galaxies could be due to mergers.

In this work, exploiting the infrared (IR) data obtained using {\it
  Herschel Space Observatory} \citep{Pilbratt2010} and 
{\it Spitzer Space Telescope} \citep{Werner2004}, 
we study the star-formation
enhancement of SFGs in close major-merger pairs since
z=1. Throughout this paper, we adopt the $\Lambda$-cosmology with $\rm
\Omega_m=0.3$ and $\rm \Omega_\Lambda = 0.7$, and $\rm H_0= 70\;
(km~sec^{-1} Mpc^{-1})$.

\section{Samples in the COSMOS Field}
Two samples of paired SFGs
with $0.2 \leq z \leq 1.0$ are studied in this work,
both are confined to ``$\rm M_*$ galaxies'' with stellar mass 
in the range of $\rm 10^{10.4} \leq M_{star} \leq 10^{11.0}\; M_\sun$.
According to \citet{Ilbert2010}, the ``$\rm M_*$''
(the turning point in the galaxy stellar
mass function) of SFGs is rather
constant against redshift since $z= 2$ : $\rm M_*\sim 10^{10.8} 
 M_\sun$ for ``intermediate active'' galaxies and
$\rm M_*\sim 10^{10.5} M_\sun$ for ``high active'' galaxies.
The ``S+S'' sample includes 124 SFGs in 62 SFG+SFG pairs 
and the ``S+E'' sample 44 SFGs in 44 mixed pairs,
all having redshifts in the range of [0.2,1.0]. 

Galaxies in both samples are taken
from the sample of close ($\rm 5 \leq r_{proj} \leq 20\;
h^{-1}\; kpc$) major-merger pairs ($\rm M_{star}^{pri}/M_{star}^{2nd} \leq 2.5$)
in the COSMOS field (CPAIR, \citealt{Xu2012}). 
The parent sample of CPAIR was
selected from the photo-$z$ catalog of the COSMOS field
\citep{Capak2007, Ilbert2009, Drory2009} with stellar
mass limits of $\rm \log (M_{min}/M_{\sun}) = [9.0, 9.4, 9.8, 10.2]$ in
four redshift bins that equally divide the redshift range of
[0.2,1.0], respectively. These limits are above the completeness 
limits for the stellar mass of the photo-$z$ sample 
\citep{Drory2009}. It was required in the
CPAIR selection that the photo-z's of the two galaxies in a pair candidate
satisfy the following criterion:
$\rm |z_{phot}^{pri}-z_{phot}^{2nd}|/(1+z_{phot}^{pri}) \leq 0.03$, where 
$z_{phot}^{pri}$ and $z_{phot}^{2nd}$ are the photo-z's of the
primary and secondary, respectively. The redshifts of both
galaxies in a pair are then assumed to be the same,
estimated by the mean of the two photo-z's.
The classification of SFGs (S) or ``passive galaxies''
(E) was based on the SED fitting and taken from \citet{Drory2009}.

Both samples are divided into 2 redshift bins: [0.2,0.6] (low-$z$ bin)
and [0.6,1.0] (high-$z$ bin). These bins are 2 times wider than those in
the original CPAIR sample. Making the size of subsamples in
redshift bins larger allows better statistics.  Since for most pairs
the beams of IR observations ($\rm FWHM > 5''$) cannot resolve a pair into
two component galaxies, the S+S sample selection was
carried out on pairs instead of on individual galaxies: a pair will be
included in the sample if $[\rm \log{(M_{star}^{pri}/M_\sun)} +
  \log{(M_{star}^{2nd}/M_\sun)}]/2$ is in the range of [10.4, 11]. The
S+E sample includes only the SFGs in the mixed pairs and excludes the
early-type galaxies.  There are 7 (11) S+S pairs (S+E pairs) in the
low-$z$ bin with the median redshift of 0.45 (0.41), and 55 (33) in the
high-$z$ bin with the median redshift of 0.81 (0.82).

The two control samples, one for each sample of paired SFGs, 
are selected from the same parent sample of CPAIR. 
In order to minimize the statistical errors in our final results
that are due to the controls,
each paired galaxy was matched by 10 control galaxies
(none being included more than once).
The following criteria were used in the selection of the controls:
(1) redshift match: $\rm |z_{phot}^{cont}-z_{phot}^{pg}|/(1+z_{phot}^{pg})
\leq 0.03$, where $z_{phot}^{cont}$ is the photo-$z$ of the control
candidate, and $z_{phot}^{pg}$ the photo-$z$ of the paired galaxy to
be matched; (2) mass match: $\rm |\log (M_{star}^{cont})-\log
(M_{star}^{pg})| \leq 0.1$, where $\rm M_{phot}^{cont}$ is the stellar
mass of the control candidate, and $\rm M_{phot}^{pg}$ the stellar mass
of the paired galaxy to be matched; (3) type match: all control
galaxies are SFGs; (4) local density match: $\rm |\rho_{nb}^{cont}
- \rho_{nb}^{pg}| \leq MAX(2, \sqrt{\rho_{nb}^{pg}})$, where $\rm
\rho_{nb}^{cont}$ is the neighbor counts 
(for neighbor galaxies of $\rm \log (M_{star}/M_\sun) \geq 10.2$)
within $\rm 1 Mpc$ (comoving) of the
control candidate, and $\rm \rho_{nb}^{pg}$ the neighbor counts around the
paired galaxy. Here, in order to have 
adequate numbers of control candidates for paired galaxies with very 
low local densities, we set a minimum ($\rm min =2$) to the density match limit
which is otherwise equal to the Poisson error of $\rm \rho_{nb}^{pg}$. 

In order to exclude interacting galaxies, 
the following two additional criteria
were applied to the selection of control candidates: 
(5) no companion within the projected distance $\rm r_{p} = 50 h^{-1} kpc$
that has  $\rm |\delta z_{phot}|/(1+z_{phot}^{cont}) \leq 0.03$ and more
massive than $\rm 0.2\times M_{star}^{cont}$; (6) $\rm A < 0.35$
and $\rm G+0.4\times A < 0.66$, where $\rm A$ is the asymmetry parameter
\citep{Conselice2000} and $\rm G$ the Gini Coefficient \citep{Lotz2008}.
The criterion (6) excludes 
morphologically disturbed galaxies that are generally associated with
mergers \citep{Lotz2008, Conselice2009}. Visual inspections of 100 
galaxies randomly picked from the control samples showed no obvious
merger candidates except for a few ambiguous cases of galaxies 
with possible weak and faint distortion features.

\section{Local Samples}
In order to 
provide the local benchmarks for the evolutionary study,
two local samples of paired SFGs, one for those in S+S pairs and another
for those in S+E pairs, and their corresponding control samples 
are also studied.
These samples are adopted from \citet{Xu2010},
who observed a nearly complete K-band selected pair sample (KPAIR)
using \textit{Spitzer} in the 3.6, 4.5, 5.8, 8, 24, 70, and 160 $\mu m$
bands. The sample was selected from cross-matches between 2MASS and
SDSS-DR3 galaxies, 
including 27 S+S and S+E pairs that have $\rm 5 \leq r_{proj} \leq 20\;
h^{-1}\; kpc$ and mass ratio of $\rm M_{star}^{pri}/M_{star}^{2nd} \leq 2.5$.
The selection criteria are nearly the same to those for the CPAIR sample.
The paired galaxies are then one-to-one matched (according to the stellar mass
and redshift) by a sample of control galaxies selected from the SWIRE survey
\citep{Lonsdale2003} and the SINGS survey \citep{Kennicutt2003}. There are
39 non-AGN SFGs in the pair sample of \citet{Xu2010}.
Among them, all paired SFGs (and their controls)
in the mass range of $\rm 10^{10.4} \leq M_{star} \leq 10^{11.0}\; M_\sun$
(stellar mass estimated from the K-band luminosity using Kroupa IMF)
are included in the local samples of this study.
There are 20 (7) SFGs in the local S+S (S+E) sample, and the same number
of galaxies in the corresponding control sample.

\tabletypesize{\footnotesize}
%\tabletypesize{\small}
\begin{deluxetable*}{ccccccccccccccccccccccccc}
%\tabletypesize{\tiny}
%\tabletypesize{\normalsize}
%\tabletypesize{\footnotesize}
%\tiny(5pt);\scriptsize(7pt);\footnotesize(8pt);\small(9pt);\normalsize(10pt)
\setlength{\tabcolsep}{0.5cm} %Tighten up the columns. See AASTeX FAQ
%\rotate
%\tablenum{1}
%\tablewidth{0pt}
%\tablecaption{The \IRAS\ RBGS LIRG and ULIRG Sample}
\tablecaption{Characteristics of the IR Data \label{tbl:IR-bands}}
%\nopagebreak
\tablehead{
 \colhead{Bands} 
& &  \colhead{Inst$\rm ^a$} 
& &  \colhead{FWHM $\rm ^b$} 
& &  \multicolumn{3}{c}{sensitivity}
& &  \multicolumn{15}{c}{detection rate ($\rm N_{detection}/N_{total}$)}
\\
\cline{7-9}
\cline{11-25}
%\\ 
& & 
& & 
& &  \colhead{$\rm f_{lim}$ $\rm ^c$}
& &  \colhead{$\rm s/\sigma$ $\rm ^d$} 
& &  \multicolumn{3}{c}{S+S pair} 
& &  \multicolumn{3}{c}{S+S contr} 
& &  \multicolumn{3}{c}{S+E pair} 
& &  \multicolumn{3}{c}{S+E contr}
\\ 
& &
& &  \colhead{(arcsec)} 
& &  \colhead{(mJy)} 
& &  
& &  \colhead{l-z$\rm ^e$} 
& &  \colhead{h-z$\rm ^f$} 
& &  \colhead{l-z} 
& &  \colhead{h-z} 
& &  \colhead{l-z} 
& &  \colhead{h-z} 
& &  \colhead{l-z} 
& &  \colhead{h-z} 
}
\startdata
 24$\mu m$  && MIPS  && 6.1  && 0.08 && 4.4 && 6/7 && 40/55 && 107/140 && 663/1110 && 9/11 &&15/33 && 90/110 && 196/330 \\
100$\mu m$  && PACS  && 7.5  &&  5.0 && 3.0 && 3/7 && 13/55 && 54/140  && 93/1110  && 5/11 && 2/33 && 54/110 && 27/330  \\
160$\mu m$  && PACS  && 11.2 && 10.2 && 3.0 && 3/7 && 12/55 && 48/140  && 84/1110  && 4/11 && 3/33 && 44/110 && 23/330  \\
250$\mu m$  && SPIRE && 18.2 && 17.4 && 3.0 && 2/7 && 11/55 && 28/140  && 56/1110  && 3/11 && 1/33 && 31/110 && 13/330  \\ 
350$\mu m$  && SPIRE && 24.9 && 18.8 && 3.0 && 3/7 &&  3/55 && 4/140   && 27/1110  && 1/11 && 1/33 && 7/110  && 4/330   \\ 
500$\mu m$  && SPIRE && 36.3 && 20.4 && 3.0 && 0/7 &&  0/55 && 1/140   && 10/1110  && 0/11 && 0/33 && 3/110  && 2/330   
%\hline
\enddata
\tablecomments{
%\begin{description}
$a)$ Insrument;
$b)$ full width at half maximum of the point spread function (PSF);
$c)$ flux limit;
$d)$ signal-to-noise ratio at the flux limit;
$e)$ low-$z$ bin ($0.2<z<0.6$);
$f)$ high-$z$ bin ($0.6<z<1$).
%\end{description}
}
\end{deluxetable*}

\section{The IR Data}
The COSMOS field was surveyed by the {\it PACS Evolutionary Probe} (PEP,
\citealt{Lutz2011}) using PACS photometer arrays 
\citep{Poglitsch2010}. The 3-$\sigma$ sensitivities 
in the PACS 100 and 160 $\mu m$ maps are
5.0 mJy and 10.2 mJy, respectively \citep{Berta2011}.
The 250, 350 and 500 $\mu m$ observations were carried
out by {\it Herschel Multi-tiered Extragalactic Survey}\footnote{http://hermes.sussex.ac.uk}
(HerMES, \citealt{Oliver2012}) using SPIRE photometer arrays
\citep{Griffin2010, Swinyard2010}.
The maps in the three SPIRE bands  \citep{Levenson2010} are
confusion limited at 3-$\sigma$ level of 17.4, 18.8, and 20.4 mJy, respectively
\citep{Nguyen2010}. In addition, we also included the \textit{Spitzer} 
24 $\mu m$ data
from the S-COSMOS survey \citep{Sanders2007}. The 
24 $\mu m$ source catalog of the COSMOS field \citep{LeFloch2009} is flux
limited at $\rm f_{24\mu m}=80\; \mu Jy$. 

We assume that 
the IR emission in an S+E pair is due to the SFG component only,
ignoring the contribution from the E component \citep{Domingue2003}.
Because most S+S pairs (median angular 
separation $\sim 3''$) are not resolved in the IR bands,
each S+S pair is treated as a single source.
Table~\ref{tbl:IR-bands} provides the main characteristics of the IR data.
In general, the
detection rates of our samples in the 5 Herschel bands are rather low.
In particular, except for S+S pairs, sources in the high-$z$ bin ($0.6<z<1$)
have detection rates $< 10\%$ in all Herschel samples. The S+S pairs 
have slightly higher detection rates 
because most of them are not resolved by Herschel, therefore each Herschel
source includes contributions from both component galaxies.
The detection rates in the MIPS 24 $\mu m$ band
are between 45.5\% and 85.7\%. 
\setcounter{table}{1}
\tabletypesize{\footnotesize}
\begin{deluxetable*}{cccccccccccccccccc}
%\tabletypesize{\tiny}
%\tabletypesize{\normalsize}
%\tabletypesize{\footnotesize}
%\tiny(5pt);\scriptsize(7pt);\footnotesize(8pt);\small(9pt);\normalsize(10pt)
%\setlength{\tabcolsep}{0.02in} %Tighten up the columns. See AASTeX FAQ
%\rotate
%\tablenum{2}
%\tablewidth{0pt}
%\tablecaption{The \IRAS\ RBGS LIRG and ULIRG Sample}
\tablecaption{IR Emission of Paired Galaxies and Control Galaxies $\rm ^a$ \label{tbl:results}}
%\nopagebreak
\tablehead{
 \colhead{sample} 
&   \colhead{$\rm N $} 
&   \colhead{$z_{med}$} 
& &  \colhead{$\rm f_{24\mu m}$} 
& &  \colhead{$\rm f_{100\mu m}$} 
& &  \colhead{$\rm f_{160\mu m}$} 
& &  \colhead{$\rm f_{250\mu m}$} 
& &  \colhead{$\rm f_{350\mu m}$} 
& &  \colhead{$\rm f_{500\mu m}$} 
& &  \colhead{$\rm \log (L_{IR})$} 
&   \colhead{$\rm \epsilon$ $\rm ^b$} 
\\
\hline
& 
& 
& &  \colhead{(mJy)} 
& &  \colhead{(mJy)} 
& &  \colhead{(mJy)} 
& &  \colhead{(mJy)} 
& &  \colhead{(mJy)} 
& &  \colhead{(mJy)} 
& &  \colhead{($\rm L_\sun$)} 
&   
}
\startdata
 S+S $\rm ^d$ &  7 &0.45  & & $\rm 0.40\pm 0.12$  & & $\rm 9.78\pm 3.99$ & & 
$\rm 14.95\pm 6.29$ & & $\rm 11.83\pm 3.89$ & & $\rm 6.72\pm 2.24$  & & 
$\rm < 3.86$ $\rm ^c$ 
& & $\rm 11.06\pm 0.13$  & $\rm 0.20\pm 0.13$ 
\\ 
         &  55 & 0.81  & & $\rm 0.13\pm 0.02$  & & $\rm 2.43\pm 0.66$ & & 
$\rm 5.00\pm 0.94$ & & $\rm 5.50\pm 0.82$ & & $\rm 4.56\pm 0.79$  & & 
$\rm 3.23\pm 0.60$
& & $\rm 11.15\pm 0.06$ & $\rm 0.04\pm 0.07$  \\ 
 Control &  140 & 0.45 & & $\rm 0.28\pm 0.03$  & & $\rm 5.75\pm 0.61$ & & 
$\rm 10.25\pm 0.91$ & & $\rm 8.75\pm 0.73$ & & $\rm 5.71\pm 0.71$  & & 
$\rm 2.58\pm 0.64$ &&
$\rm 10.86\pm 0.04$  & \\ 
         &  1100 & 0.82  & & $\rm 0.13\pm 0.01$  & & $\rm 2.03\pm 0.19$ & & 
$\rm 4.13\pm 0.24$ & & $\rm 5.16\pm 0.27$ & & $\rm 4.32\pm 0.23$  & & 
$\rm 2.34\pm 0.21$ &&
$\rm 11.11\pm 0.03$  & \\ 
\hline
%& & & & & & & & & &  & & & & & & &  \\
 S+E &  11 & 0.41  & & $\rm 0.35\pm 0.08$  & & $\rm 8.69\pm 2.63$ & & 
$\rm 13.01\pm 4.47$ & & $\rm 13.18\pm 3.74$ & & $\rm 8.43\pm 2.93$  & & 
$\rm < 6.15 $ $\rm ^c$ 
& & $\rm 10.96\pm 0.11$  & $\rm 0.05\pm 0.11$  \\ 
         &  33 & 0.82  & & $\rm 0.14\pm 0.03$  & & $\rm 1.76\pm 1.07$ & & 
$\rm 4.35\pm 1.85$ & & $\rm 5.95\pm 1.65$ & & $\rm < 3.27$ $\rm ^c$ & & 
$\rm < 3.55$ $\rm ^c$
& & $\rm 11.13\pm 0.12$  & $\rm 0.04\pm 0.12$  \\ 
 Control &  110 & 0.41  & & $\rm 0.39\pm 0.04$  & & $\rm 8.19\pm 1.07$ & & 
$\rm 14.06\pm 1.38$ & & $\rm 11.45\pm 1.02$ & & $\rm 6.18\pm 0.75$  & & 
$\rm 2.51\pm 0.69$ 
& & $\rm 10.90\pm 0.04$ & \\ 
         &  330 & 0.82  & & $\rm 0.14\pm 0.01$  & & $\rm 1.95\pm 0.22$ & & 
$\rm 3.40\pm 0.37$ & & $\rm 5.13\pm 0.43$ & & $\rm 4.26\pm 0.46$  & & 
$\rm 2.66\pm 0.41$ 
& & $\rm 11.09\pm 0.04$  &
%\hline
\enddata
\tablecomments{
%\begin{description}
$a)$ Mean IR fluxes were estimated through stacking, and the errors through
bootstrapping;
$b)$ mean sSFR enhancement index;
$c)$ 3-$\sigma$ upper-limit ($\rm = f_{lim}/\sqrt{N}$, where $\rm f_{lim}$
is the 3-$\sigma$ sensitivity limit of the band, and N the size of the sample);
$d)$ mean IR fluxes, errors and upper-limits for paired SFGs were derived by dividing corresponding values of S+S pairs by 2.}
\end{deluxetable*}

\section{Stacking Analyses and Results}
Because of the low IR detection rates, we studied the IR emission of
our sources through stacking. Two stacking methods are
considered. The first is a modified version of ``cleaned
stacking''. In its original form \citep{Zheng2007}, there are four
steps in this method: (1) detect and extract all bright sources above
a given detection threshold from the observational image; 
(2) separate the target sources into detected and undetected by
matching them with the catalog of detected sources; (3) stack the
images of undetected sources, which are cut from the residual map 
with the detected sources extracted, and
measure the flux; (4) derive 
the mean flux of target sources $\rm f
= (\sum_i f_{det,i}+f_{stack})/N_{total}$, 
where $\rm f_{det.i}$ is the flux of the i-th detected source,\
$\rm f_{stack}$ the flux measured in the stacked image, and $\rm N_{total}$ 
the total number of sources in the target sample. 
Compared to uncleaned stacking \citep{Dole2006}, cleaned stacking
suffers less confusion problems due to unrelated background sources.
However, the result might be biased due to flux boosting for
detected sources, in particular if the detection limit is set at a low
$\rm s/\sigma$ level. In order to avoid this bias, we 
made the following modifications. First, the detection thresholds were set
relatively high (at $\rm s/\sigma \sim 6$) when 
doing the detection and extraction
on observed images using StarFinder \citep{Diolaiti2000}. Then,
the extractions corresponding to detected target sources were restored back to 
the residual map; the subsequent stacking, using the 
residual map with detected background sources extracted, 
went through all target sources
including both detected and undetected. Finally, the mean flux of a given
band for a given sample was derived by $\rm f = f_{stack}/N_{total}$. 

Some sources in our samples are marginally resolved
in the IR images, in particular the S+S pairs with $\rm sep > 5''$
($\sim 7\%$ of all S+S pairs). In order to ensure that a
stacked image includes all fluxes from all
sources, we searched in
the catalog of detected sources
for all matches to a given S+S pair using a searching radius of
$\rm 0.5\times \sqrt{(FWHM^2+sep^2)}$.  For SFGs in the S+E sample
and two control samples, which are generally point sources in all IR
bands (e.g. the mean Petrosian radius of galaxies in the S+S control
sample is $\rm 1.2''$ with a dispersion of $0.6''$), the searching
radius is $\rm 0.5\times FWHM $.  In the photometry of the stacked
images, we choose to use a constant aperture of radius $\rm R =
11.2''$ for the three MIPS and PEP bands at 24, 100 and 160 $\mu
m$. The corresponding aperture corrections are 1.09, 1.05, 1.21,
respectively, according to our empirical tests using StarFinder.
\begin{figure*}[!htb]
\epsscale{1.1}
\plottwo{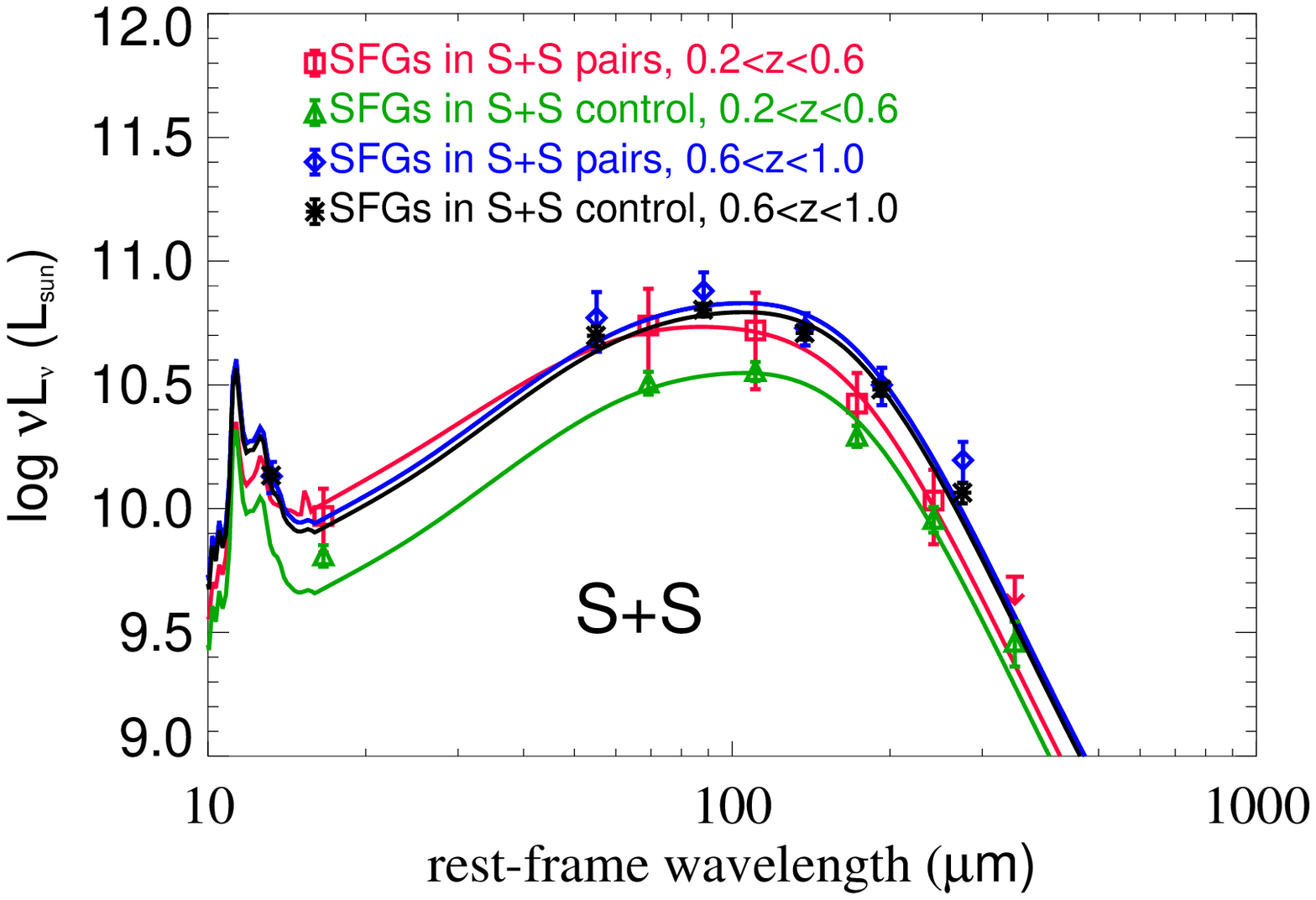}{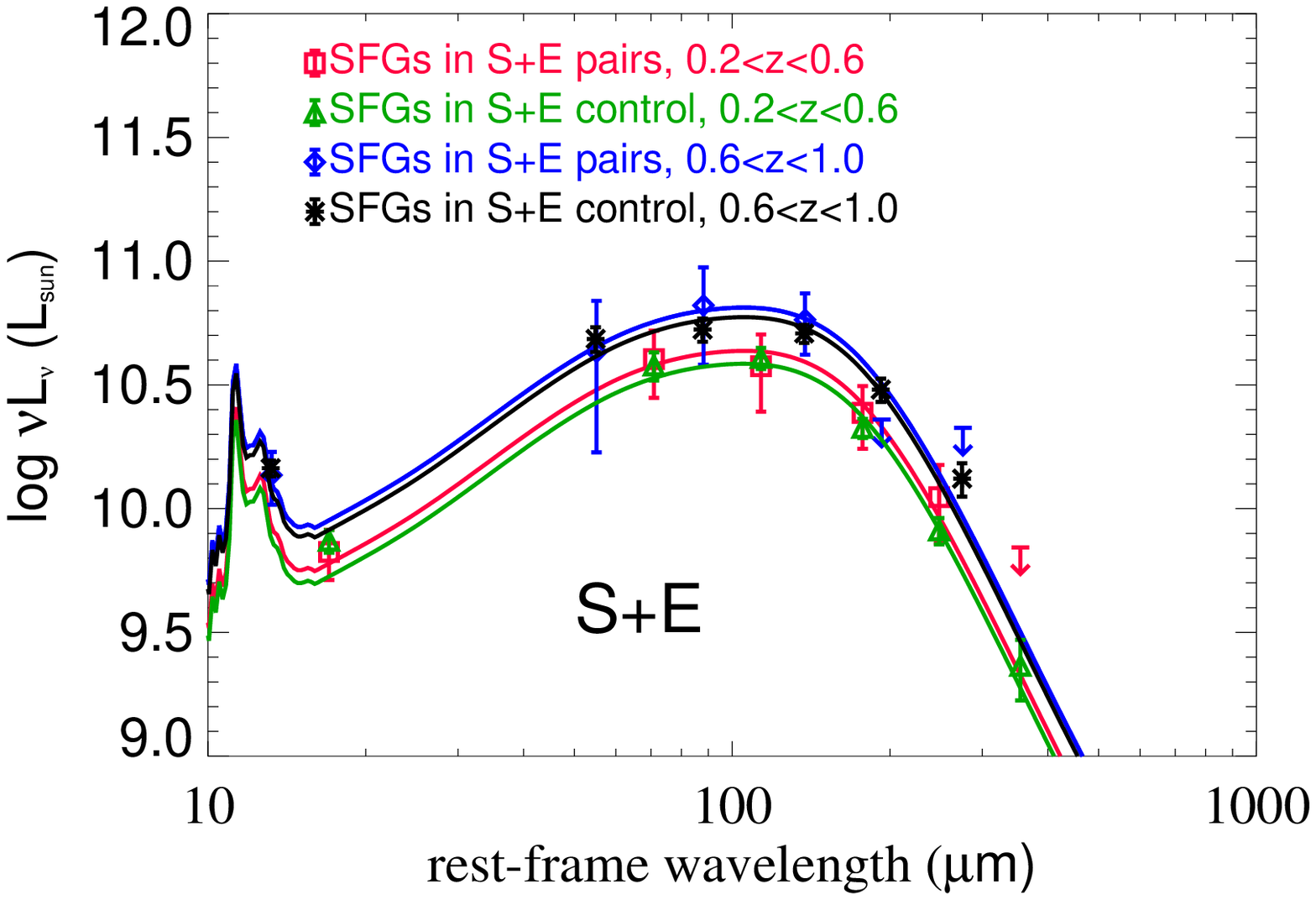}
\caption{{\bf Left Panel:} The mean IR SED of SFGs in the
S+S sample and of those in the S+S control sample. The red curve is
the best fit to data of SFGs in S+S pairs with $0.2<z<0.6$,
using the semi-empirical SED of Arp~244 (the Antennae Galaxies) 
taken from~ \citet{Xu2001}. The other three curves are
the best fits to data of SFGs in S+S pairs with $0.6<z<1.0$ and
SFGs in the S+S control sample
 with $0.2<z<0.6$ and $0.6<z<1.0$, respectively,
using the same semi-empirical SED of normal Sc galaxy NGC~6181 \citep{Xu2001}.
{\bf Right Panel:} The mean IR SED of SFGs in the
S+E sample and of those in the S+E control sample. All curves are
the best fits to data 
using the same semi-empirical SED of normal Sc galaxy NGC~6181.
}
\label{fig:seds}
\end{figure*}

In the second method, the ``covariance'' method
\citep{Marsden2009}, the mean flux of a sample of
target sources is estimated by $\rm f = \sum_k D_k/N$, where $\rm D_k$
is the measured flux density (in units of Jy/beam) in the map pixel
that contains the k-th source in the target sample. The map,
without any cleaning of unrelated
sources, is background subtracted with zero mean.  The method explicitly
assumes that target sources are randomly distributed
in the sky according to the Poisson statistics, 
which is certainly not valid for paired galaxies, therefore 
it should not be applied to individual galaxies in S+S pairs.
Also, this method may significantly under-estimate the mean 
flux of target sources if some of them are resolved (even only partially).

Our experiments showed that for SPIRE images at 250, 350 and 500 $\mu m$, 
which are predominantly confusion limited \citep{Oliver2012}, the
mean fluxes derived using the covariance method and cleaned stacking
are all consistent with each other within 1$\sigma$ error; and results
of the covariance method generally have slightly lower errors than
those of cleaned stacking. On the other hand, in the MIPS 24  $\mu m$  
and PACS 100 $\mu m$ bands, the mean fluxes of S+S pairs derived 
using the covariance method are systematically lower than those obtained using
cleaned stacking, presumably due to missing flux of resolved sources
in the former. Therefore, in the final analyses, we chose to use the cleaned
stacking method for the 24, 100 and 160 $\mu m$ bands,
and the covariance method for the 250, 350 and 500 $\mu m$ bands.

The results are listed in Table~\ref{tbl:results}.
All errors were estimated by bootstrapping \citep{Efron1979}. 
For a given sample and band,
the error includes both the statistical dispersion
of the sample and the measurement noise (instrumental and confusion noise).
A 3-$\sigma$ upper-limit is assigned to a mean flux if it 
is less than $\rm f_{lim}/\sqrt{N}$, where $\rm f_{lim}$
is the 3-$\sigma$ sensitivity limit of the band, and N the size of the sample.
For the S+S sample, N is the number of pairs; 
for other samples, N is the number of galaxies.  
For SFGs in the S+S sample, the mean fluxes, 
errors, and upper-limits 
were all obtained by dividing corresponding values
of the pairs by a factor of 2.  

We neglected the bias due to clustering of IR sources
\citep{Bethermin2012}.  Our experiments using
simulations based on the algorithm described in \citet{Bethermin2012}
showed that for our (mass-selected) samples the clustering bias is not
strong, $\sim 15\%$ for the mean fluxes in the 500 $\mu m$ band
and $\sim 7\%$ for those in other bands, without
significant dependence on redshift.  Because of the local density
match between the control samples and the pair samples (Section 2),
the biases for pairs and for controls should be the same and therefore
cancel each other out in the star-formation enhancement
analysis. On the same ground, any other possible systematic biases in the
mean fluxes (e.g. due to filtering of the maps) have also been neglected.

\section{Star-Formation Enhancement in Paired SFGs}
Using the median redshift of each subsample, the mean IR fluxes obtained
by stacking are converted to the rest-frame $\rm \nu L_{\nu}$ and plotted
in Fig.~\ref{fig:seds}. The semi-empirical SEDs in the
SED library of \citet{Xu2001} were searched for the best fits to
these data. The SED library\footnote{
http://spider.ipac.caltech.edu/staff/cxu/sed\_lib/intro\_sed\_lib.html}
is based on the IRAS, ISO, and ground based
sub-mm observations of 837 IRAS 25 $\mu m$ band selected galaxies.
Among bright FIR sources ($\rm f_{100\mu m} > 10$ Jy) in the library, we found 
that the SED of local merger Arp~244 (the Antennae Galaxies)
fits best the data of SFGs in S+S sample in
the low-$z$ bin ($0.2<z<0.6$). Interestingly,
the mean SEDs of all other subsamples, including 
the high-$z$ bin ($0.6<z<1$) of the S+S sample and
both redshift bins of the S+E 
sample, have similar shapes and can be best fitted by
the SED of the same local normal Sc galaxy NGC~6181 (Fig.~\ref{fig:seds}).

The inverse variance weighted
least-squares fit to data in the 100, 160, 250, 350 $\mu m$ bands
(i.e. data close to the IR peak), using
the corresponding best-fit SED, was carried out for each subsample to determine
the mean total IR luminosity $\rm L_{IR} (\hbox{5 -- 1000}\mu m)$. 
The variance of $\rm L_{IR}$ was estimated by
\begin{equation} 
\rm \sigma^2 = \sum_i \sum_j {Cov_{i,j}\over \sigma_i \sigma_j} /
                \left(\sum_i 1/\sigma_i^2 \right)^2,
\end{equation} 
where all summations are over the four bands at 
100, 160, 250, 350 $\mu m$, and $\rm Cov_{i,j}$ is the covariance
between band $i$ and band $j$. 
If the errors in different bands
were independent with each other, i.e. $\rm Cov_{i,j} = 0$ for $i \neq j$,
the variance of $\rm L_{IR}$ could have been estimated by
$\rm \sigma^2 = 1/\sum_i  1/\sigma_i^2$. However for our data we found
rather significant covariance between different bands
(also estimated using bootstrapping). 

\begin{figure}[!htb]
\epsscale{1.3}
\plotone{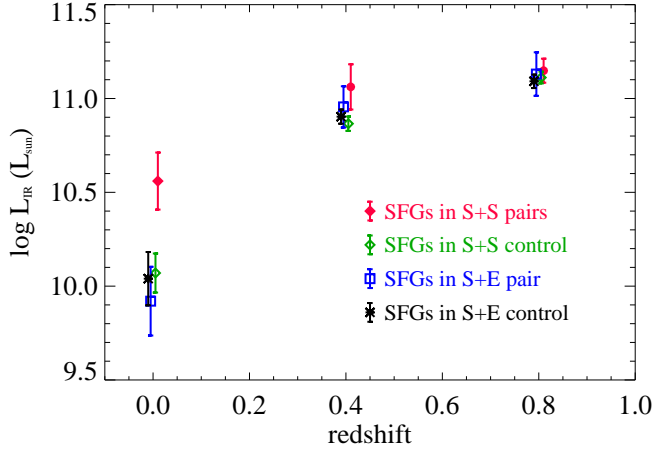}
\caption{Plot of mean $\rm \log(L_{IR})$ vs. redshift. Each data point
and the error bar at $z>0$ were derived through fitting the
 corresponding SED in Fig.~\ref{fig:seds}. Data points for the
local ($z=0$) samples were obtained using Spitzer observations 
of \citet{Xu2010}. There are 20 and 7 SFGs in the local S+S and S+E
samples. Their control samples have the same sizes.
}
\label{fig:lir}
\end{figure}

\begin{figure}[!htb]
\epsscale{1.3}
\plotone{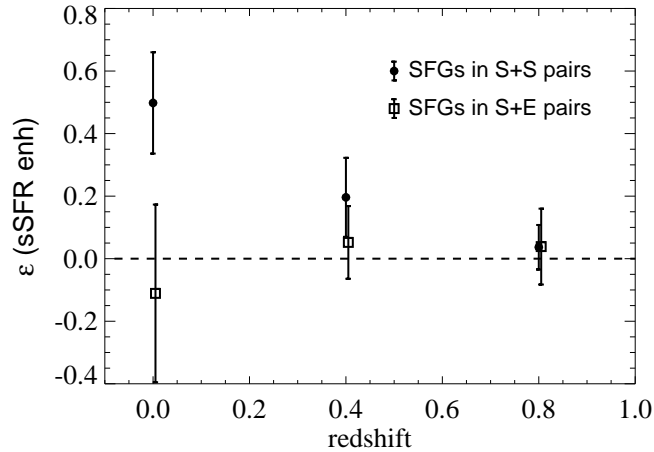}
\caption{Plot of mean sSFR enhancement index,
$\rm \epsilon = \log(sSFR_{pair})-\log(sSFR_{cont})$, vs. redshift.
Because each pair sample and its
control sample have the same mean stellar mass, $\rm \epsilon =
\log(L_{IR,pair})-\log(L_{IR,cont})$.
}
\label{fig:enhancement}
\end{figure}

Resulting $\rm L_{IR}$'s are listed in 
Table~\ref{tbl:results} and plotted in Fig.~\ref{fig:lir}
against redshift. Also plotted are mean IR luminosities of
SFGs in the local samples.
The mean $\rm L_{IR}$'s of the local
samples are derived using \textit{Spitzer} 
data at 8, 24, 70, and 160 $\mu m$
\citep{Xu2010}. Fig.~\ref{fig:enhancement} shows the sSFR enhancement of
SFGs in S+S pairs and in S+E pairs.  The sSFR enhancement index is defined
as 
\begin{eqnarray}
\rm \epsilon & = & \log(sSFR_{pair})-\log(sSFR_{cont}) \\
             & = & \log(L_{IR,pair})-\log(L_{IR,cont}). 
\end{eqnarray}
The last equation is valid because,
by design, each pair sample and its
control sample have the same mean stellar mass. 
Also, we neglected the uncertainties of the $\rm L_{IR}$  
as an SFR indicator that are due either to
the dust heating by old stars or to the UV radiation
of young stars escaping the dust absorption.
These uncertainties should not affect the
calculation of $\rm \epsilon$ because the effects on
the paired SFGs and on the controls should be the same
and therefore should cancel each other out.

For SFGs in S+S pairs, 
data are consistent with un-enhanced star-formation  
in paired galaxies in the high-$z$ bin ($0.6<z<1$). 
They show an average sSFR enhancement index
of $\rm \epsilon=0.04\pm 0.07$, and their mean SED
and that of their controls are both
best-fitted by the same SED of a normal disk galaxy (NGC~6181).
In the low-$z$ bin ($0.2<z<0.6$), mainly because of the small
number of S+S pairs in the bin,
the results are less clear-cut: their 
average sSFR enhancement index is $\rm \epsilon=0.20\pm 0.13$,
indicating a weak enhancement at rather low significance level of 87\%
(i.e. at 1.5-$\sigma$ level). The mean
SED is best fitted by that of a galaxy merger
(Arp~244) but, given the large error bars of individual 
data points, a fit with the SED of the normal spiral NGC~6181 cannot 
be ruled out. 
For SFGs in S+E pairs in both the low-$z$ and high-$z$ bins, results on both
the sSFR enhancement and the SED comparison 
are consistent with no enhancement in their star-formation activities.

\section{Discussion}
\subsection{S+S Pairs: Less Star-Formation Enhancement at Higher $z$}
The non-enhancement of the star-formation in 
SFGs in S+S pairs with $z\sim 0.8$ and the rather weak 
star-formation enhancement in those
with $z\sim 0.4$, in contrast with the strong and significant
star-formation enhancement ($\rm \epsilon=0.50\pm 0.16$)
in the local sample (Fig.~\ref{fig:enhancement}), reveal
a clear trend for the star-formation enhancement to decrease with redshift.
It is unlikely that this trend is due to an increase of
the spurious pairs fraction (SPF) with redshift. The probability for
selecting (unphysical) projected pairs is significantly reduced by
CPAIR selection criteria $\rm 5 \leq r_{proj} \leq 20\; h^{-1}\; kpc$
and $\rm M_{star}^{pri}/M_{star}^{2nd} \leq 2.5$, confining the companion
search to a very small sky area ($\rm \sim 10\; arcsce^2$) and a
narrow mass range. Indeed, as shown by the Monte Carlo simulations
\citep{Xu2012}, the spurious pair fraction 
due to projected pairs (given the photo-z
error of the COSMOS survey and the pair selection criterion $
\delta z_{phot}/(1+z_{phot}) \leq 0.03$) is only $7\pm 3$\% at z=0.3
and $9\pm 3$\% at $z=0.9$.  Even after including the effect of galaxy
clustering and counting pairs with $\rm \delta V > 500\; km\; s^{-1}$ as spurious
pairs, the spurious pair fraction 
is still rather low in the CPAIR sample: $\rm SPF =
0.20\pm 0.06$ at $z=0.3$ and $\rm SPF = 0.22\pm 0.06$ at $z=0.9$. The
local ($z=0$) pairs were selected based on spectroscopic redshifts
\citep{Xu2010}. For them, $\rm SPF = 0.06\pm 0.05$ \citep{Xu2012}, due
to unbound galaxies that are clustered together (therefore
having recession velocity difference as low as $\rm
\delta V \leq 500\; km\; s^{-1}$). Statistically, the effect of
the contamination of spurious pairs can be corrected as following
\begin{equation}
\rm \epsilon_{true} =  \epsilon_{obs}/(1-SPF).
\end{equation}
For SFGs in S+S pairs in the high-$z$ bin ($0.6<z<1$), this
correction results in a small change in the mean enhancement index,
from $\rm \epsilon_{obs} = 0.04$ to $\rm \epsilon_{true} = 0.05$.
In summary, the low spurious pair fractions in our pair samples 
shall not affect the results on the sSFR enhancement significantly.

In parallel to the decline of the sSFR enhancement in S+S pairs, mean
$\rm L_{IR}$ of normal SFGs in the control sample increases rapidly
with redshift (Fig.~\ref{fig:lir}).  Strong SFR evolution in normal
disk galaxies since $z\sim 1$ has been well documented
\citep{Bell2005, Zheng2007, Elbaz2007, Noeske2007a}, 
and has been attributed to the
increase of gas fraction with redshift in these galaxies. Indeed, CO
observations of massive SFGs at $z\sim 1$ show significantly higher
molecular gas content compared to local disk galaxies
\citep{Tacconi2010}.  According to \citet{Hopkins2009}, the ratio
between mass of gas consumed by an interaction-induced nuclear starburst
and the disk gas mass decreases with increasing disk gas fraction as $\rm
f_{burst}/f_{gas}\propto (1-f_{gas})$, where $\rm f_{burst}$ is the gas
fraction (among the total baryonic mass) consumed by the starburst and
$\rm f_{gas}$ the disk gas fraction. This is because the
gravitational torque imposed by the stellar disk to the gas disk is
less effective when $\rm f_{gas}$ is high, therefore less disk gas can
sink to the nuclear region by losing angular momentum to stars.  We
argue that the negative cosmic evolution of the sSFR enhancement in
S+S pairs is due to the increase of $\rm f_{gas}$ with redshift.

The sSFR enhancement in local ($z=0$) close major-merger pairs has been
well established \citep{Kennicutt1987, Xu1991, Barton2000, Nikolic2004,
  Ellison2008}.  Meanwhile, \citet{Xu2010} found that the enhancement
only occurs in massive SFGs ($\rm \gsim 10^{10} M_\sun$) in S+S
pairs. Paired SFGs with low $\rm M_{star}$ have in general higher $\rm
f_{gas}$ than massive SFGs and show no significant sSFR
enhancement. It is worth noting that studies of B-band selected
interacting galaxies, which are biased for low mass SFGs, found no or
very weak sSFR enhancement \citep{Bergvall2003, Knapen2009}. It is
plausible that the high $\rm f_{gas}$ ($\gsim 30\%$) plays a
significant role in the non-enhancement of the sSFR in low mass
paired SFGs.

There have only been a few previous statistical studies of the SFR of
interacting galaxies in the universe at $z>0.1$.  Using UV and optical
SFR indicators, which are prone to dust attenuation, both
\citet{LeFevre2000} and \citet{Bridge2010} found that the SFR of
interacting galaxies at $z \lsim 1$ are enhanced by a factor of $\sim
2$ compared to the field galaxies, and the SFR enhancement increases
with the redshift. On the other hand, \citet{LeFevre2000} noticed a
difference between ``upcoming major-mergers'' and ``ongoing
major-mergers''; the former do not show significant SFR
enhancement (as measured by [OII] equivalent width).
\citet{Bridge2010} also found that, in their sample of morphologically
selected mergers, early stage mergers (close pairs with tidal bridges)
have SFRs similar to field galaxies. These are consistent with our
results of the non-enhancement for paired SFG at $z=0.8$, because our
pair selection favors early stage mergers. However, different from
early stage mergers in \citet{LeFevre2000} and \citet{Bridge2010},
SFGs in S+S pairs at $z=0$ do show significant sSFR enhancement
in this study.  Given that our pairs at different redshifts are selected
using the same selection criteria, it is unlikely that the ratio
between the early stage and late stage mergers changes with the
redshift. 
\citet{Lin2007}, comparing the median $\rm L_{IR}/M_{star}$ ($\rm L_{IR}$ being
derived from the 24$\mu m$ flux) of a
sample of close pairs and that of pseudo-pairs
constructed using conrol galaxies, found an
enhancement of a factor of $\sim 2$ for pairs in the redshift range
$0.1 < z <1$. However, by assigning a very low $\rm L_{IR}$ 
value ($\rm 10^{9} L_\sun$) to undetected control
galaxies, \citet{Lin2007} may have under-estimated the 
median  $\rm L_{IR}/M_{star}$ of control pairs and over-estimated the
enhancement of the close pairs (most being unresolved in the 24$\mu m$ survey). 
Our result on the mean 24$\mu m$
emission of S+S pairs in the high-$z$ bin differs from that of 
\citet{Lin2007}, showing no enhancement compared to the controls
(Table~\ref{tbl:results}). 
\citet{Jogee2009} studied the SFR of morphologically selected mergers
with $0.24<z<0.80$, using both the UV and MIR (24 $\mu m$) data,
and found modest enhancement compared to normal galaxies.
Studying an IR selected sample observed by Herschel, \citet{Hwang2011}
found a factor of 1.8 -- 4.0 sSFR enhancement for galaxies in close
SFG-SFG pairs with redshifts between 0 and 1.2, with little variation in 
the enhancement versus redshift.  However, given the blending
effect, the true enhancement indicated by results of \citet{Hwang2011}
is significantly less (by a factor of $\sim 0.5$). Also
their results are consistent with a decreasing star
formation enhancement at higher $z$ because
higher $z$ sources are likely to be affected more severely by
blending. 
 
\subsection{Cosmic Evolution of Extreme Starbursts}
According to simulations \citep{Scudder2012}, 
samples of close pairs ($\rm 5 \leq r_{proj} \leq 20\;
h^{-1}\; kpc$) include mostly merging galaxies undergoing the first
close encounter and those just before the final coalescence,
both stages lasting $\sim 10^8$ years. They miss final stage mergers
(with the time scale of $\sim 10^7$ years)
that are too close to be identified as binaries, and those already coalesced.
Early IRAS observations \citep{Kennicutt1987, Sanders1996}
already revealed that final stage mergers show much stronger star-formation
enhancement than paired galaxies, and extreme starbursts such as ULIRGs
occur almost exclusively within these final stage mergers.
Given the selection effects for pair samples, the cosmic evolution
of these extreme starbursts is not probed by the data in this study.

On the other hand, there are indications in the literature that
properties of extreme starbursts in final stage mergers might also
change with redshift. \citet{Rujopakarn2011} found that ULIRGs at
$z\sim 1$ and sub-millimeter galaxies (SMGs) at $z\sim 2$ have much
more extended star-formation distributions than local ULIRGs, and
argued that these are isolated galaxies with high $\rm \Sigma_{SFR}$.
However, studying the morphology of 70 $\mu m$ sources in the S-COSMOS
survey, \citet{Kartaltepe2010} concluded that major mergers dominate
the ULIRG population at $z\lsim 1$ .  The CO and radio continuum
observations of \citet{Biggs2008} and \citet{Engel2010} demonstrate
that SMGs are also dominated by major mergers even though, different
from local ULIRGs that typically have sizes of $\lsim 1$ kpc
\citep{Rujopakarn2011}, the extreme starbursts in high-$z$ mergers are
much more extended ($\sim 5$ kpc). These authors suggested that the
difference between the size of local ULIRGs and that of high-$z$ SMGs is
due to the selection effect in the sense that SMGs are preferentially
earlier stage mergers than local ULIRGs. However, it is possible that
the star-formation in high-$z$ SMGs (and high-$z$ ULIRGs) is less
centrally peaked because the strength of the nuclear starburst is
suppressed due to the higher gas fraction in the disk, according to
the same physical mechanism \citep{Hopkins2009} that we invoked in the
interpretation for the weakening of the sSFR enhancement in higher z
S+S pairs.  It will be interesting to check in future studies whether
the percentage of extreme starbursts (with sSFR enhancement $\gsim$ a
factor of 100) among major mergers also decreases with redshift.

\subsection{SFGs in S+E Pairs: No sSFR Enhancement}
For SFGs in S+E pairs, the sSFR enhancement index $\rm
\epsilon$ is consistent with 0 at all redshifts. This agrees with the
\textit{Spitzer} results of \citet{Xu2010} which showed no sSFR enhancement for
SFGs in S+E pairs in any stellar mass bin in a local close
major-merger pair sample. \citet{Hwang2011} also found that SFGs with
close early-type neighbors are not sSFR enhanced.  The difference
between SFGs in S+S pairs and in S+E pairs indicates that, in addition to
gravitational tidal effects, the sSFR in a paired galaxy is
influenced by the immediate surrounding environment.  This hypothesis
is in agreement with the correlation between sSFR's of the primaries
and secondaries in major-merger S+S pairs (i.e. the ``Holmberg effect'',
\citealt{Kennicutt1987, Xu2010}). On the other hand, \citet{Xu2010}
found no significant difference between the local densities around S+S
and S+E pairs in their local pair sample, which is confirmed again here
for pairs with $z=0.2$ -- 1.0; the average counts of neighbors of $\rm
M_{star} \geq 10^{10.2} M_\sun$ within 1 Mpc projected distance and
with $\delta z_{phot}/(1+z_{phot}) \leq 0.007$ (the 1$\sigma$
error of photo-z, \citealt{Ilbert2009}) are $\rm 3.69\pm 0.45$ and
$\rm 3.73\pm 0.52$ for S+S pairs and S+E pairs,
respectively. Therefore, the linear scale of the environment effect
must be less than 1 Mpc. Because most galaxies in close major-merger
pairs have entered the virial radius of the companions a long time ago,
the two galaxies in a pair are likely to share the same IGM gas in a
common dark matter halo (DMH).  The non-enhancement of
the sSFR in SFGs in S+E pairs and the ``Holmberg effect'' for S+S
pairs suggest a significant role of the IGM within a DMH to the
sSFR of galaxies residing in the DMH.  For example, when a DMH has
strong (weak) ``cold flows'' \citep{Dekel2009, Keres2009},
galaxies inside it may have abundant (scarce) cold gas supply to fuel 
active star-formation. This hypotheses seems particularly attractive because
it can explain both the  ``Holmberg effect'' for S+S pairs and
the non-enhancement of the SFG in S+E pairs.  
Another scenario for the IGM modulation of
the sSFR involves the hot IGM gas which can strip the cold ISM gas
in embedded galaxies \citep{Park2005}, though this may have difficulty
in explaining the ``Holmberg effect'' for S+S pairs because 
detections of the hot IGM gas around spiral galaxies are very rare
\citep{Bensen2000, Anderson2011}.

\vskip1truecm
\noindent{\it Acknowledgments}:
Dr Xianzhong Zheng is thanked for kindly providing 
the software and instructions for the ``clean stacking''.
We acknowledge support from the Science and Technology Facilities Council  (grant number ST/I000976/1).
PACS has been developed by a consortium of institutes led by MPE (Germany) and including UVIE (Austria); KU Leuven, CSL, IMEC (Belgium); CEA, LAM (France); MPIA (Germany); INAF-IFSI/OAA/OAP/OAT, LENS, SISSA (Italy); IAC (Spain). This development has been supported by the funding agencies BMVIT (Austria), ESA-PRODEX (Belgium), CEA/CNES (France), DLR (Germany), ASI/INAF (Italy), and CICYT/MCYT (Spain).
SPIRE has been developed by a consortium of institutes led by Cardiff Univ. (UK) and including: Univ. Lethbridge (Canada); NAOC (China); CEA, LAM (France); IFSI, Univ. Padua (Italy); IAC (Spain); Stockholm Observatory (Sweden); Imperial College London, RAL, UCL-MSSL, UKATC, Univ. Sussex (UK); and Caltech, JPL, NHSC, Univ. Colorado (USA). This development has been supported by national funding agencies: CSA (Canada); NAOC (China); CEA, CNES, CNRS (France); ASI (Italy); MCINN (Spain); SNSB (Sweden); STFC, UKSA (UK); and NASA (USA).
The data presented in this paper will be released through the \textit{Herschel} Database in Marseille (HeDaM, \url{http://hedam.oamp.fr/HerMES}).

%\vskip3truecm

\bibliographystyle{apj}
\bibliography{/Volumes/Seagate/data1/bibliography/ckxu_biblio}

\begin{thebibliography}{77}
\expandafter\ifx\csname natexlab\endcsname\relax\def\natexlab#1{#1}\fi

\bibitem[{Abraham {et~al.}(1996)Abraham, Tanvir, Santiago,
  {et~al.}}]{Abraham1996}
Abraham, R., Tanvir, N.~K., Santiago, B.~X., {et~al.} 1996, MNRAS, 279

\bibitem[{Anderson \& Bregman(2011)}]{Anderson2011}
Anderson, M.~E. \& Bregman, J. 2011, ApJ, 737, 22

\bibitem[{Barton {et~al.}(2000)Barton, Geller, \& Kenyon}]{Barton2000}
Barton, E.~J., Geller, M.~J., \& Kenyon, S.~J. 2000, ApJ, 530, 660

\bibitem[{Baugh {et~al.}(2005)}]{Baugh2005}
Baugh, C.~M. {et~al.} 2005, MNRAS, 359, 119

\bibitem[{Bell {et~al.}(2005)Bell, Papovich, Wolf, {et~al.}}]{Bell2005}
Bell, E.~F., Papovich, C., Wolf, C., {et~al.} 2005, ApJ, 625, 23

\bibitem[{Bensen {et~al.}(2000)Bensen, Bower, Frenk, \& White}]{Bensen2000}
Bensen, A.~J., Bower, B.~G., Frenk, C.~S., \& White, S.~D.~M. 2000, MNRAS, 314,
  555

\bibitem[{Bergvall {et~al.}(2003)Bergvall, Laurikainen, \&
  Aalto}]{Bergvall2003}
Bergvall, N., Laurikainen, E., \& Aalto, S. 2003, A\&A, 405, 31

\bibitem[{Berta {et~al.}(2011)Berta, Magnelli, Nordon, {et~al.}}]{Berta2011}
Berta, S., Magnelli, B., Nordon, R., {et~al.} 2011, A\&A, 532, 49

\bibitem[{B\'{e}thermin {et~al.}(2012)B\'{e}thermin, LeFloc$'$h, Ilbert,
  {et~al.}}]{Bethermin2012}
B\'{e}thermin, M., LeFloc$'$h, E., Ilbert, O., {et~al.} 2012

\bibitem[{Biggs \& Ivison(2008)}]{Biggs2008}
Biggs, A.~D. \& Ivison, R.~J. 2008, MNRAS, 385, 893

\bibitem[{Bridge {et~al.}(2007)Bridge, Appleton, Conselice,
  {et~al.}}]{Bridge2007}
Bridge, C.~R., Appleton, P.~N., Conselice, C.~J., {et~al.} 2007, ApJ, 659, 931

\bibitem[{Bridge {et~al.}(2010)Bridge, Carlberg, \& Sullivan}]{Bridge2010}
Bridge, C.~R., Carlberg, R.~G., \& Sullivan, M. 2010, ApJ, 709, 1067

\bibitem[{Brinchmann {et~al.}(1998)Brinchmann, Abraham, Shade,
  {et~al.}}]{Brinchmann1998}
Brinchmann, J., Abraham, R., Shade, D., {et~al.} 1998, ApJ, 499, 112

\bibitem[{Capak {et~al.}(2007)}]{Capak2007}
Capak, P. {et~al.} 2007, ApJS, 172, 284

\bibitem[{Conselice {et~al.}(2000)Conselice, Bershady, \&
  Jangren}]{Conselice2000}
Conselice, C.~J., Bershady, M.~A., \& Jangren, A. 2000, ApJ, 529, 886

\bibitem[{Conselice {et~al.}(2003)Conselice, Bershady,
  {et~al.}}]{Conselice2003}
Conselice, C.~J., Bershady, M.~A., {et~al.} 2003, AJ, 126, 1183

\bibitem[{Conselice {et~al.}(2009)Conselice, Yang, C., \&
  Bluck}]{Conselice2009}
Conselice, C.~J., Yang, C., C., \& Bluck, A.~F.~L. 2009, MNRAS, 394, 1956

\bibitem[{Daddi {et~al.}(2007)Daddi, Dickinson, Morrison, {et~al.}}]{Daddi2007}
Daddi, E., Dickinson, M., Morrison, G., {et~al.} 2007, ApJ, 670, 156

\bibitem[{Dekel {et~al.}(2009)}]{Dekel2009}
Dekel, A. {et~al.} 2009, Nature, 475, 451

\bibitem[{Diolaiti {et~al.}(2000)}]{Diolaiti2000}
Diolaiti, E. {et~al.} 2000, SPIE, 4007, 879

\bibitem[{Dole {et~al.}(2006)Dole, Lagache, Puget, {et~al.}}]{Dole2006}
Dole, H., Lagache, G., Puget, J.-L., {et~al.} 2006, A\&A, 451, 417

\bibitem[{Domingue {et~al.}(2003)Domingue, Sulentic, Xu,
  {et~al.}}]{Domingue2003}
Domingue, D.~L., Sulentic, J.~W., Xu, C., {et~al.} 2003, AJ, 125, 555

\bibitem[{Driver {et~al.}(1995)Driver, Windhorst, Ostrander,
  {et~al.}}]{Driver1995}
Driver, S., Windhorst, R., Ostrander, E.~J., {et~al.} 1995, ApJL, 449

\bibitem[{Drory {et~al.}(2009)}]{Drory2009}
Drory, N. {et~al.} 2009, ApJ, 707, 1995

\bibitem[{Efron(1979)}]{Efron1979}
Efron, B. 1979, The Annals of Statistics, 7, 1

\bibitem[{Elbaz {et~al.}(2007)Elbaz, Daddi, Borgne, {et~al.}}]{Elbaz2007}
Elbaz, D., Daddi, E., Borgne, D.~L., {et~al.} 2007, A\&A, 468, 33

\bibitem[{Elbaz {et~al.}(2011)Elbaz, Dickinson, Hwang, {et~al.}}]{Elbaz2011}
Elbaz, D., Dickinson, M., Hwang, H.~S., {et~al.} 2011, A\&A, 533, 119

\bibitem[{Ellison {et~al.}(2008)Ellison, Patton, Simard,
  {et~al.}}]{Ellison2008}
Ellison, S.~L., Patton, D.~R., Simard, L., {et~al.} 2008, AJ, 135, 1877

\bibitem[{Engel {et~al.}(2010)Engel, Taconi, I.Davis, {et~al.}}]{Engel2010}
Engel, H., Taconi, L.~L., I.Davis, R., {et~al.} 2010, ApJ, 724, 233

\bibitem[{Glazebrook {et~al.}(1995)Glazebrook, Ellis, Santiago, \&
  Grifﬁths}]{Glazebrook1995}
Glazebrook, K., Ellis, R., Santiago, B., \& Grifﬁths, R. 1995, MNRAS, 275

\bibitem[{Griffin {et~al.}(2010)}]{Griffin2010}
Griffin, M. {et~al.} 2010, A\&A, 518

\bibitem[{Guiderdoni {et~al.}(1998)}]{Guiderdoni1998}
Guiderdoni, B. {et~al.} 1998, MNRAS, 295, 877

\bibitem[{Hopkins {et~al.}(2009)}]{Hopkins2009}
Hopkins, P.~F. {et~al.} 2009, ApJ, 691, 1186

\bibitem[{Hwang {et~al.}(2011)Hwang, Elbaz, Dickinson, {et~al.}}]{Hwang2011}
Hwang, H.~S., Elbaz, D., Dickinson, M., {et~al.} 2011, A\&A, 535, 60

\bibitem[{Ilbert {et~al.}(2009)}]{Ilbert2009}
Ilbert, O. {et~al.} 2009, ApJ, 690, 1236

\bibitem[{Ilbert {et~al.}(2010)}]{Ilbert2010}
---. 2010, ApJ, 709, 644

\bibitem[{Jogee {et~al.}(2009)Jogee, Miller, Penner, {et~al.}}]{Jogee2009}
Jogee, S., Miller, S.~H., Penner, K., {et~al.} 2009, ApJ, 697, 1971

\bibitem[{Kartaltepe {et~al.}(2007)}]{Kartaltepe2007}
Kartaltepe, J.~S. {et~al.} 2007, ApJS, 172, 320

\bibitem[{Kartaltepe {et~al.}(2010)}]{Kartaltepe2010}
---. 2010, ApJ, 721, 98

\bibitem[{Kennicutt {et~al.}(1987)Kennicutt, Keel, van~der Hulst,
  {et~al.}}]{Kennicutt1987}
Kennicutt, R.~C., Keel, W., van~der Hulst, J., {et~al.} 1987, AJ, 93, 1001

\bibitem[{Kennicutt {et~al.}(2003)}]{Kennicutt2003}
Kennicutt, R.~C. {et~al.} 2003, PASP, 115, 928

\bibitem[{Keres {et~al.}(2009)}]{Keres2009}
Keres, D. {et~al.} 2009, MNRAS, 309, 160

\bibitem[{Knapen \& James(2009)}]{Knapen2009}
Knapen, J.~H. \& James, P. 2009, ApJ, 698, 1437

\bibitem[{LeF\'{e}vre {et~al.}(2000)LeF\'{e}vre, Abraham, \&
  Lilly}]{LeFevre2000}
LeF\'{e}vre, O., Abraham, R., \& Lilly, S.~J. 2000, MNRAS, 311, 565

\bibitem[{LeFloc$'$h {et~al.}(2009)LeFloc$'$h, Aussel, Ilbert,
  {et~al.}}]{LeFloch2009}
LeFloc$'$h, E., Aussel, H., Ilbert, O., {et~al.} 2009, ApJ, 703, 222

\bibitem[{Levenson {et~al.}(2010)Levenson, Marsden, Zemvov,
  {et~al.}}]{Levenson2010}
Levenson, L., Marsden, G., Zemvov, M., {et~al.} 2010, ApJ, 709, 97

\bibitem[{Lin {et~al.}(2007)Lin, Koo, Weiner, {et~al.}}]{Lin2007}
Lin, L., Koo, D.~C., Weiner, B.~J., {et~al.} 2007, ApJL, 660, 51

\bibitem[{Lin {et~al.}(2008)Lin, Patton, \& Koo}]{Lin2008}
Lin, L., Patton, D.~R., \& Koo, D.~C. 2008, ApJ, 681, 232

\bibitem[{Lonsdale {et~al.}(2003)}]{Lonsdale2003}
Lonsdale, C.~J. {et~al.} 2003, PASP, 115, 897

\bibitem[{Lotz {et~al.}(2008)Lotz, Davis, Faber, {et~al.}}]{Lotz2008}
Lotz, J.~M., Davis, M., Faber, S.~M., {et~al.} 2008, ApJ, 672, 177

\bibitem[{Lutz {et~al.}(2011)}]{Lutz2011}
Lutz, D. {et~al.} 2011, A\&A, 532, 90

\bibitem[{Marsden {et~al.}(2009)Marsden, Ade, Bock, {et~al.}}]{Marsden2009}
Marsden, G., Ade, P.~A., Bock, J.~J., {et~al.} 2009, ApJ, 707, 1729

\bibitem[{Nguyen {et~al.}(2010)}]{Nguyen2010}
Nguyen, H.~T. {et~al.} 2010, A\&A, 518

\bibitem[{Nikolic {et~al.}(2004)Nikolic, Cullen, \& Alexander}]{Nikolic2004}
Nikolic, B., Cullen, H., \& Alexander, P. 2004, MNRAS, 355, 874

\bibitem[{Noeske {et~al.}(2007)Noeske, Weiner, Faber, {et~al.}}]{Noeske2007a}
Noeske, K.~G., Weiner, B.~J., Faber, S.~M., {et~al.} 2007, ApJL, 660, 43

\bibitem[{Oliver {et~al.}(2012)}]{Oliver2012}
Oliver, S. {et~al.} 2012, arXiv: 1203.2562

\bibitem[{Pannella {et~al.}(2009)Pannella, Carilli, Daddi,
  {et~al.}}]{Pannella2009}
Pannella, M., Carilli, C.~L., Daddi, E., {et~al.} 2009, ApJ, 698, 116

\bibitem[{Park \& Choi(2005)}]{Park2005}
Park, C. \& Choi, Y.-Y. 2005, ApJ, 635

\bibitem[{Peng {et~al.}(2010)}]{Peng2010}
Peng, Y. {et~al.} 2010, ApJ, 721, 193

\bibitem[{Pilbratt {et~al.}(2010)}]{Pilbratt2010}
Pilbratt, G.~L. {et~al.} 2010, A\&A, 518

\bibitem[{Poglitsch {et~al.}(2010)}]{Poglitsch2010}
Poglitsch, A. {et~al.} 2010, A\&A, 518

\bibitem[{Rodighiero {et~al.}(2011)Rodighiero, Daddi, Baronchelli,
  {et~al.}}]{Rodighiero2011}
Rodighiero, G., Daddi, E., Baronchelli, I., {et~al.} 2011, ApJ, 739, 40

\bibitem[{Rujopakarn {et~al.}(2011)Rujopakarn, Rieke, Eisenstein, \&
  Juneau}]{Rujopakarn2011}
Rujopakarn, W., Rieke, G.~H., Eisenstein, D.~I., \& Juneau, S. 2011, ApJ, 726,
  93

\bibitem[{Sanders \& Mirabel(1996)}]{Sanders1996}
Sanders, D.~B. \& Mirabel, I.~F. 1996, ARA\&A, 34, 749

\bibitem[{Sanders {et~al.}(2007)}]{Sanders2007}
Sanders, D.~B. {et~al.} 2007, ApJS, 172

\bibitem[{Scudder {et~al.}(2012)}]{Scudder2012}
Scudder, J.~M. {et~al.} 2012, arXiv: 1207.4791

\bibitem[{Shi {et~al.}(2009)Shi, Rieke, Lotz, {et~al.}}]{Shi2009}
Shi, Y., Rieke, G., Lotz, J., {et~al.} 2009, ApJ, 697, 1764

\bibitem[{Somerville {et~al.}(2000)}]{Somerville2001}
Somerville, R.~S. {et~al.} 2000, MNRAS, 320, 504

\bibitem[{Swinyard {et~al.}(2010)}]{Swinyard2010}
Swinyard, B. {et~al.} 2010, A\&A, 518

\bibitem[{Tacconi {et~al.}(2010)}]{Tacconi2010}
Tacconi, L.~J. {et~al.} 2010, Nature, 463, 387

\bibitem[{Werner {et~al.}(2004)}]{Werner2004}
Werner, M.~W. {et~al.} 2004, ApJS, 154, 1

\bibitem[{Wuyts {et~al.}(2011)Wuyts, Forster, Natascha, {et~al.}}]{Wuyts2011}
Wuyts, S., Forster, S., Natascha, M., {et~al.} 2011, ApJ, 738, 106

\bibitem[{{Xu} {et~al.}(2001){Xu}, {Lonsdale}, {Shupe}, {O'Linger}, \&
  {Masci}}]{Xu2001}
{Xu}, C., {Lonsdale}, C.~J., {Shupe}, D.~L., {O'Linger}, J., \& {Masci}, F.
  2001, \apj, 562, 179

\bibitem[{{Xu} \& {Sulentic}(1991)}]{Xu1991}
{Xu}, C. \& {Sulentic}, J.~W. 1991, \apj, 374, 407

\bibitem[{{Xu} {et~al.}(2010){Xu}, {Domingue}, {Cheng}, {et~al.}}]{Xu2010}
{Xu}, C.~K., {Domingue}, D., {Cheng}, Y., {et~al.} 2010, \apj, 713, 330

\bibitem[{Xu {et~al.}(2012)Xu, Zhao, Scoville, {et~al.}}]{Xu2012}
Xu, C.~K., Zhao, Y., Scoville, N., {et~al.} 2012, ApJ, 747, 85

\bibitem[{Zheng {et~al.}(2007)Zheng, Bell, Papovich, {et~al.}}]{Zheng2007}
Zheng, X.~Z., Bell, E.~F., Papovich, C., {et~al.} 2007, ApJL, 661, 41

\end{thebibliography}
%\bibliography{ckxu_biblio}

%\clearpage
%%%%%%%%%%%%%%%%%%%%%%%%%%%%%%%%%%%%%%%%%%%%%%%%%%%%%%%%%%%%%%%%%
% The references section.
%%%%%%%%%%%%%%%%%%%%%%%%%%%%%%%%%%%%%%%%%%%%%%%%%%%%%%%%%%%%%%%%%
%\end{landscape}

\end{document}